\documentclass[11pt,a4paper]{article}
\usepackage[utf8]{inputenc}
\usepackage{jinstpub}
\usepackage{url}
\usepackage{booktabs}
\usepackage{upgreek}
\usepackage{graphicx}
\usepackage{subcaption}
\usepackage{lineno}
\newcommand{\degC}[1]{\ensuremath{#1}\,\textdegree{}C}

\newcommand{\ntestedtotal}{368}

\title{Improved quality tests of R11410-21 photomultiplier tubes for the XENONnT experiment}

\note{Now at Physikalisches Institut, Universit\"at Freiburg, 79104 Freiburg, Germany.} %1
\note{Corresponding author.}
\note{Now at Department of Physics and Chemistry, University of L'Aquila, 67100 L'Aquila, Italy and INFN-Laboratori Nazionali del Gran Sasso and Gran Sasso Science Institute, 67100 L'Aquila, Italy.}
\note{Now at Institute for Advanced Research and Kobayashi-Maskawa Institute for the Origin of Particles and the Universe, Nagoya University, Nagoya 464–8602, Japan.}

\author[a]{V.~C.~Antochi,}
\author[b]{L.~Baudis,}
\author[c,1]{J.~Bollig,}
\author[b,1,2]{A.~Brown,}
\author[d]{R.~Budnik,}
\author[c]{D.~Cichon,}
\author[a]{J.~Conrad,}
\author[a,3]{A.~D.~Ferella,}
\author[b]{M.~Galloway,}
\author[c,2]{L.~Hoetzsch,}
\author[b,4]{S.~Kazama,}
\author[d]{G.~Koltman,}
\author[d]{H.~Landsman,}
\author[c]{M.~Lindner,}
\author[a,2]{J.~Mahlstedt,}
\author[c]{T.~Marrod\'an~Undagoitia,}
\author[a]{B.~Pelssers,}
\author[b]{G.~Volta,}
\author[c]{O.~Wack,}
\author[b]{J.~Wulf}
\affiliation[a]{Oskar Klein Centre, Department of Physics, Stockholm University, AlbaNova, Stockholm SE-10691, Sweden}
\affiliation[b]{Physik-Institut, Universit\"at Z\"urich, 8057  Z\"urich, Switzerland}
\affiliation[c]{Max-Planck-Institut f\"ur Kernphysik, 69117 Heidelberg, Germany}
\affiliation[d]{Department of Particle Physics and Astrophysics, Weizmann Institute of Science, Rehovot 7610001, Israel}

\emailAdd{abrown@physik.uzh.ch}
\emailAdd{luisa.hoetzsch@mpi-hd.mpg.de}
\emailAdd{joern.mahlstedt@fysik.su.se}

\abstract{
Photomultiplier tubes (PMTs) are often used in low-background particle physics experiments, which rely on an excellent response to single-photon signals and stable long-term operation.
In particular, the Hamamatsu R11410 model is the light sensor of choice for liquid xenon dark matter experiments, including XENONnT.
The same PMT model was also used for the predecessor, XENON1T, where issues affecting its long-term operation were observed.
Here, we report on an improved PMT testing procedure which ensures optimal performance in XENONnT.
Using both new and upgraded facilities, we tested 368 new PMTs in a cryogenic xenon environment.
We developed new tests targeted at the detection of light emission and the degradation of the PMT vacuum through small leaks, which can lead to spurious signals known as afterpulses, both of which were observed in XENON1T.

We exclude the use of 26 of the 368 tested PMTs and categorise the remainder according to their performance.
Given that we have improved the testing procedure, yet we rejected fewer PMTs, we expect significantly better PMT performance in XENONnT.
}


\begin{document}
\maketitle
\flushbottom

\section{Introduction}
\label{sec:intro}

Dual-phase xenon time projection chambers (TPCs) perform the most sensitive direct searches for dark matter today.
They rely on photosensors which are able to detect xenon's 175 nm~\cite{Fujii:2015aa} vacuum ultraviolet (VUV) scintillation light while operated at cryogenic temperatures.
Despite being a relatively old technology, photomultiplier tubes (PMTs) are still the first choice of photon detector for dark matter searches.
In particular, the Hamamatsu R11410 is a 3-inch PMT used in three leading dark matter experiments, XENON~\cite{Aprile:2017aty}, PandaX~\cite{Cao:2014jsa}, and LUX-ZEPLIN~\cite{Mount:2017qzi}, as well as other experiments including NEXT~\cite{Monrabal:2018xlr} and RED~\cite{Akimov:2012aya}.
The R11410's low dark count rate and high quantum efficiency at 175\,nm are essential characteristics for experiments which depend upon reliably detecting the few photons produced in low-energy interactions.
Its low-resistance bialkali photocathode~\cite{Nakamura:2010zzk} provides excellent performance at low temperature.
Finally, the materials used for the variant employed in the XENONnT TPC (R11410-21) were carefully selected based on an extensive radioactivity screening campaign. This resulted in a very low level of radioactivity per PMT, with a $^{238}$U activity of less than 13\,mBq and $^{228}$Th activity of $0.4 \pm 0.1$\,mBq~\cite{Aprile:2015lha}. This is crucial when searching for extremely rare signals.

Despite their many strengths, individual PMTs can suffer from specific problems. These include the random emission of small amounts of light at the level of individual photons, which contributes to the experimental background.
In addition, vacuum degradation, resulting in secondary signals known as afterpulses, can worsen a detector's event reconstruction precision.
In this article, we describe the characterisation of \ntestedtotal{} new Hamamatsu R11410-21 PMTs for XENONnT~\cite{Aprile:2020vtw}, delivered between April 2017 and December 2019.
XENONnT is an upgrade of XENON1T, with almost three times as much xenon (5.9\,t) in its target and requires 246 more PMTs, bringing the total to 494.
While both experiments use the same PMT model~\cite{Barrow:2016doe}, in the new production cycle Hamamatsu took measures to mitigate light emission by preventing photons produced inside the PMTs from escaping and affecting other PMTs.

Studies of different variants of the R11410 PMT are reported in~\cite{Lung:2012pi,Baudis:2013xva,Paredes:2018hxp}, while \cite{Barrow:2016doe} describes the tests performed on the R11410-21 PMTs used in the XENON1T experiment, 178 of which are used in XENONnT.

In contrast to the XENON1T testing campaign, all new PMTs -- instead of just a subset of 44 tubes -- were tested both in liquid and gaseous xenon.
To enable this large-scale liquid-xenon testing regime, we used an upgraded facility at the University of Zurich as well as a newly constructed test facility at Stockholm University.
In addition, a subset of PMTs were tested in gaseous argon at the Max-Planck-Institut für Kernphysik (MPIK) in Heidelberg.
All PMTs were monitored for at least two weeks to detect excessive light emission and afterpulsing, both of which can prohibit their long term operation in XENONnT.
We introduced a new test simulating the high illumination experienced by PMTs during calibration campaigns. This test is focused on intermittent light emission, which can be triggered by such high illumination.
All PMTs' amplification gains were measured, as well as a subset's single-photon and timing resolution.
On the basis of these tests, we determine that 342 of the 368 PMTs perform satisfactorily and are suitable for use in XENONnT.
We rank these 342 using a point-based system, discussed in section \ref{sec:discuss}.
This forms the basis for selecting and arranging them within the detector.

\section{Overview of testing facilities and procedure}
\label{sec:facilities}

A similar testing procedure was followed at all three facilities, which are summarised in table \ref{tab:facilities}.
We tested a total of \ntestedtotal{} PMTs in liquid and gaseous xenon to verify their long-term operational stability. A subset of 161 were also tested at MPIK for additional characterisation.

\begin{table}[h!]
    \centering
    \begin{tabular}{l  c  c  c c} \toprule
         & \textbf{Zurich} & \textbf{Stockholm} & \multicolumn{2}{c}{\textbf{MPIK}}  \\ \cmidrule{4-5}
        &&& Cryogenic & Faraday cage \\
        \midrule
        Environment & \multicolumn{2}{c}{Liquid and gaseous xenon} & Gaseous argon & Air \\
        Temperature &  \degC{-100}  &  \degC{-100}  &  \degC{-100}  &  Ambient \\
        Cycle duration & 2 weeks & 2 weeks & 1 day & 2--3 days \\
        PMTs per cycle & 10 & 14 & 12 & 12 \\
        Tested PMTs & 105 & 260 & 161 & 161\\ \bottomrule
    \end{tabular}
    \caption{Overview of the three facilities used to test the 368 PMTs. Of the 161 PMTs tested at MPIK, all were subsequently tested in Stockholm apart from three, which had already been disqualified on the basis of the MPIK measurements.}
    \label{tab:facilities}
\end{table}

\subsection{Xenon testing facilities}
The design and operation of the two xenon facilities, at the University of Zurich and Stockholm University, are very similar.
A vacuum-insulated cryostat, housing the PMTs being tested, forms the core of each facility.
The cryostat is filled with either liquid or cold gaseous xenon, making it possible to test the PMTs in similar conditions to those they will encounter in XENONnT~\cite{Aprile:2020vtw}.

\begin{figure}[ht]
\centering
\begin{subfigure}{.33\textwidth}
\includegraphics[width=\textwidth]{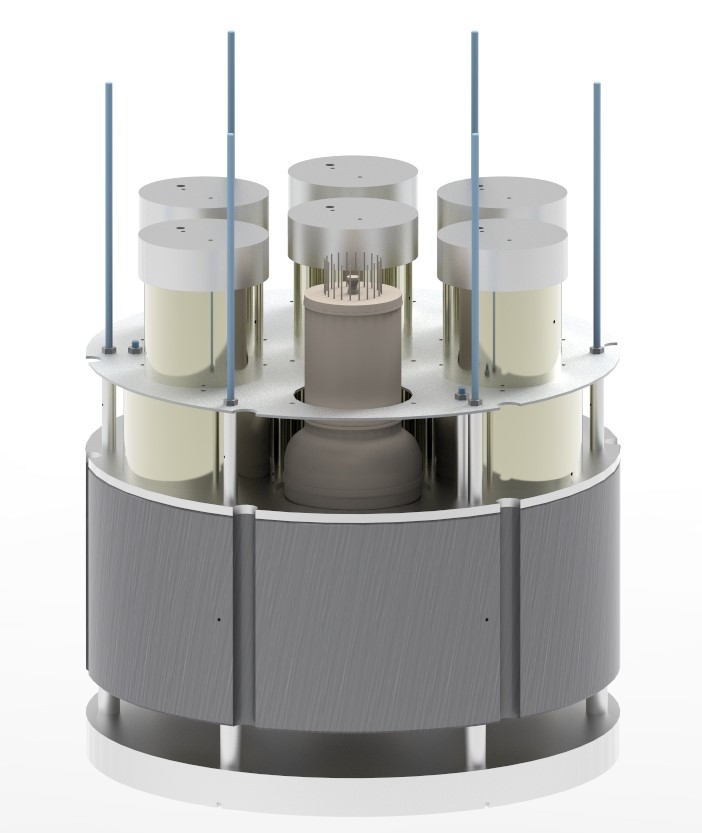}
\end{subfigure}
\hspace{1cm}
\centering
\begin{subfigure}{0.25479579207\textwidth}
\includegraphics[width=\textwidth]{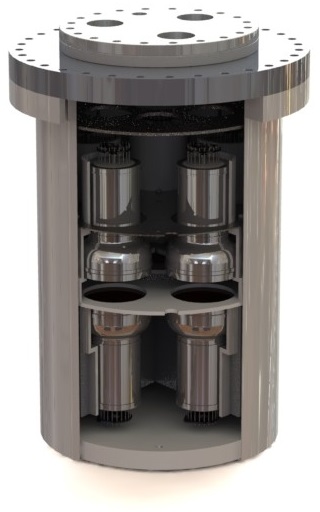}
\end{subfigure}
\caption{Illustration of PMTs in their holding structure in the Stockholm (left) and Zurich (right) facilities.}
\label{fig:xenon_facilities}
\end{figure}

In Stockholm, the cryostat holds fourteen PMTs, arranged in two layers of seven facing one another window-to-window, separated by a few millimetres (figure~\ref{fig:xenon_facilities}, left).
The facility in Zurich, which was already used for the XENON1T testing campaign~\cite{Baudis:2013xva}, was upgraded to double its capacity.
It can thus hold ten PMTs, arranged in two layers of five, as shown in figure \ref{fig:xenon_facilities} (right).
As we discuss in section~\ref{sec:le}, this arrangement with pairs of facing PMTs makes it possible to measure the rate of light being emitted by each PMT.

Both facilities are cooled by a pulse-tube refrigerator, enabling consistent and precise temperature regulation.
Blue LEDs inside the cryostat illuminate the PMTs with $\sim$470\,nm wavelength light, which is within their sensitive regime.
These are used in conjunction with an external pulse generator to provide a consistent, reproducible illumination which can be used to characterise PMTs' gains and afterpulses.
The PMTs are read out in parallel by CAEN v1724 analogue-to-digital converters; this model is also used by the XENONnT experiment.

The PMTs were cooled down at least twice, for a minimum duration of one week each time.
During the first week, the PMT testing chamber was filled with liquid xenon such that the PMTs were completely covered.
During the second week, the PMTs were held in a gaseous xenon atmosphere at its boiling point (178\,K at 2\,bar).
In both cases, the absolute pressure was maintained at around 2\,bar, similar to that used in dark matter experiments.
While cold, a variety of data were taken to characterise the PMTs' performance, including measurements of their gain, afterpulse spectra, dark count rates and light emission.

\subsection{General characterisation and argon testing facility}

The testing facility at the MPIK was already used in the XENON1T PMT testing campaign~\cite{Barrow:2016doe} and consists of two setups.
The first is housed in a light-tight 30\,m$^3$ steel box acting as a Faraday cage.
This provides a low-noise environment for simultaneous precision measurements of twelve PMTs at room temperature.
Each PMT slot is equipped with an optical fibre connected to a custom-made LED board providing 1.4\,ns long light pulses at a wavelength of 380\,nm.
The setup allows us to determine the PMTs' dark count rate, gain, single-photon resolution, timing performance and afterpulse characteristics.

The second setup consists of two circular PMT support structures, holding 12 PMTs in total. We upgraded the structure with six optical fibres, allowing high-intensity LED illumination of the PMTs. Similarly to the two xenon facilities, the structure is placed inside a cryostat, with the PMTs facing each other \mbox{window-to-window}.
The cryostat can be filled with gas that is cooled to liquid xenon temperature of about \degC{-100} using a copper cooling coil flushed with liquid nitrogen. For this testing campaign, argon gas was used instead of the nitrogen gas which was used for the XENON1T tests in order to improve the leak-detection sensitivity.

The two setups share a common data acquisition system~\cite{Bauer:2011ne, Barrow:2016doe}.
PMTs' gain and single-photon resolution are measured using a charge-to-digital converter.
A time-to-digital converter coupled to a discriminator provides precise measurements of electron transit times and afterpulse timing, and a scaler module is used to determine their dark count rates.

All PMTs tested at the MPIK facility undergo the same measurement procedure in batches of 12.
General characterisation measurements are performed in the Faraday cage setup to reduce the impact of external noise.
In the cryogenic setup, the PMTs are cooled to \degC{-100} in a 1\,bar gaseous argon atmosphere. During the cold period of about five hours, the PMT behaviour is studied using measurements of the dark count rate and tests for light emission. 
All PMTs undergo this procedure at least twice, in order to be sensitive to changes resulting from thermal cycling. Afterpulse spectra before and after the two thermal cycles are measured in the Faraday cage setup to check for vacuum degradation.

\section{PMT performance}
\label{sec:tests}

Our evaluation of the \ntestedtotal{} new PMTs' performance is presented in this section. We describe the PMTs' general characteristics, including their quantum efficiency, amplification behaviour and timing properties, in section~\ref{sec:generalchars}. These results are summarised in table~\ref{tab:generalchars}. Our tests related to light emission are presented in section~\ref{sec:le}. Finally, section~\ref{sec:ap} focuses on studies involving PMT afterpulses.

\begin{table}[]
    \centering
    \begin{tabular}{l  c  c  c} \toprule
        \textbf{Property} & \textbf{Unit} & \multicolumn{2}{c}{\textbf{Result}}  \\ \cmidrule{3-4}
         & & Mean & Spread \\
        \midrule
        Quantum efficiency & \% & 34.0 & 2.8 \\
        Gain at 1.5\,kV & $10^6$ & 8.4 & 2.3 \\
        HV for gain of $5\times10^6$ & kV & 1.41 & 0.04 \\
        SPE resolution (median) & \% & 25.1 & 1.5 \\
        Peak-to-valley ratio (median) & - & 4.3 & 0.4 \\
        Transit time spread & ns & 9.2 & 0.5 \\ \bottomrule
    \end{tabular}
    \caption{Summary of general PMT characteristics. For each property, the mean and standard deviation across all PMTs tested are reported. For the SPE resolution and the peak-to-valley ratio the PMTs were operated with a gain between $4\times10^{6}$ and $6\times10^{6}$. Section~\ref{sec:generalchars} contains details of these measurements.}
    \label{tab:generalchars}
\end{table}

\subsection{General characteristics}
\label{sec:generalchars}

%\subsubsection{Quantum efficiency}

\begin{figure}[b]
\centering
\includegraphics{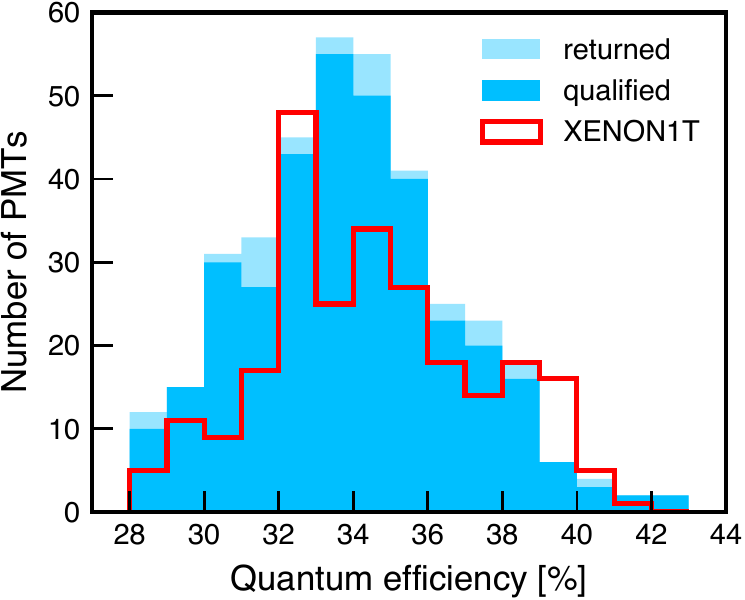}
\caption{Quantum efficiency (QE) distribution for the full set of the \ntestedtotal{} tested PMTs. The darker shaded blue distribution corresponds to the PMTs that passed all tests described in this report and thus qualified for use in XENONnT. The lighter shaded blue distribution stacked on top represents PMTs that failed any of the tests and were returned to the manufacturer for replacement. The QE distribution of the 248 PMTs used in the XENON1T detector are shown in red~\cite{Barrow:2016doe}. All values are provided by Hamamatsu.}
\label{fig:qe}
\end{figure}

The quantum efficiency (QE) of a PMT is defined as the ratio between the number of photoelectrons (PEs) emitted from the photocathode and the number of incident photons. It is measured for each PMT by the manufacturer at a temperature of \degC{25} for 175\,nm photon wavelength. We show the QE distribution for all \ntestedtotal{} tested PMTs in comparison to the XENON1T PMTs in figure \ref{fig:qe}.
The lower end of the distributions are truncated due to the contractually agreed minimum QE of 28\%. Both the distributions of all PMTs and the subset of qualified tubes have a mean QE of 34.0\%, with a standard deviation of 2.8\%. Some PMTs reach exceptionally high QEs of more than 40\%. These values are very similar to those of the XENON1T PMTs, with a mean and standard deviation of 34.5\% and 2.8\%\:\cite{Barrow:2016doe}.

The QE of the R11410 PMT model for 175\,nm photons has been shown to have a relative increase of 10\% to 15\% at \degC{-100}~\cite{Lyashenko:2014rda}. This is an effect of photoelectrons being lost after scattering off lattice phonons in the cathode: at lower temperatures there are fewer phonons so fewer photoelectrons are lost.
Due to the manufacturer’s definition of QE, the values reported here include the effect of double photoelectron emission (DPE), where one incident photon releases two photoelectrons from the photocathode. Several publications report measurements of the DPE probability for the R11410 PMT model, with values of around 22\,\% at both room and liquid xenon temperature~\cite{Faham:2015kqa,Paredes:2018hxp}.
Similarly, the best-fit value for the DPE probability extracted from the XENON1T detector response model is 21.9\,\%~\cite{Aprile:2019dme}.
Correcting for a 22\,\% DPE probability and a 10\,\% relative increase at cryogenic temperature, the mean QE of the \ntestedtotal{} tested PMTs becomes~30.7\,\%.

%\subsubsection{Amplification gain and resolution}

The amplification behaviour of PMTs is determined by their gain, defined as the multiplication factor between the number of photoelectrons emitted from the cathode and the number of electrons in the PMT signal output at the end of the dynode chain. The gain can be extracted from charge spectra acquired by illuminating the PMTs with single photons.
For all tested PMTs, we measure the gains for a set of supply voltages ranging from 1.2\,kV to 1.7\,kV. Here we report the gain at~1.5\,kV as well as the supply voltage corresponding to a gain of $5\times10^6$. It is determined by fitting the gain as a function of supply voltage, using their power law dependence.
The PMT gains are calculated either by using the model independent method described in~\cite{Saldanha:2016mkn} or by fitting the single photoelectron (SPE) charge spectra~\cite{Barrow:2016doe}.
The functional form of the fit includes a Gaussian noise component, single and multiple PE Gaussian peaks, and an exponential drop-off between the noise component and the SPE peak. The latter accounts for under-amplified SPE pulses.
\begin{figure}[b]
    \includegraphics{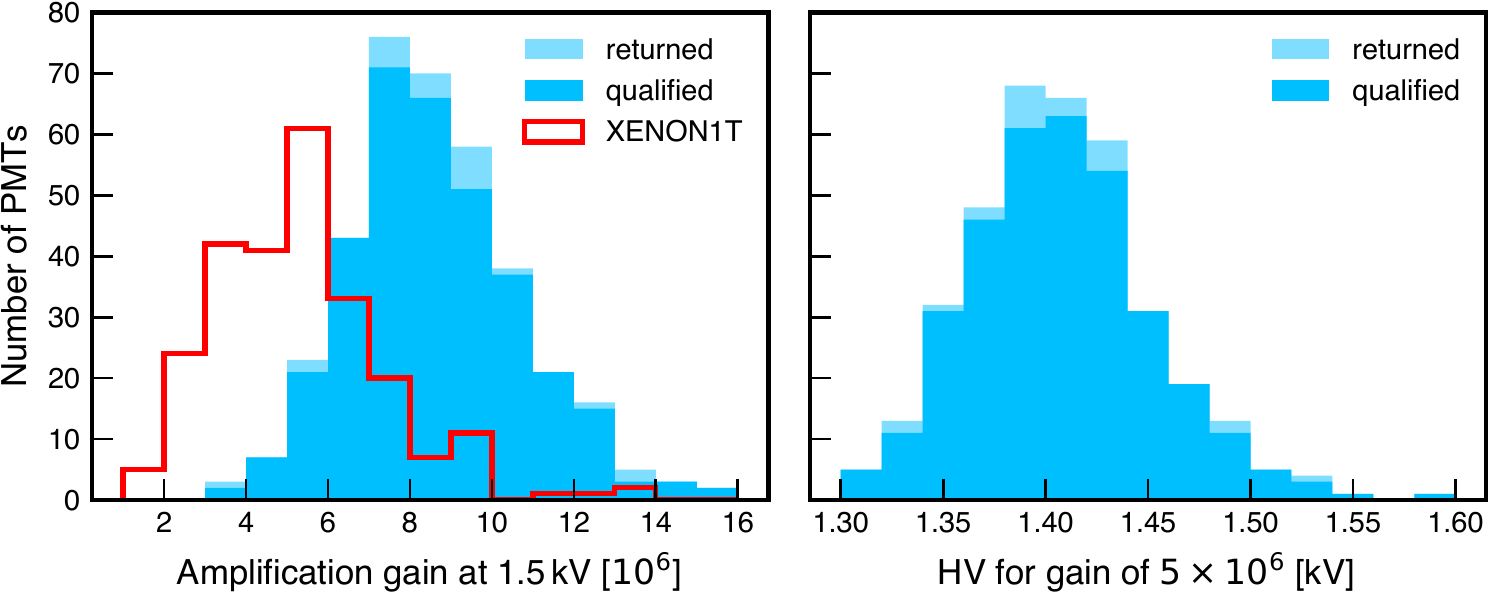}
    
    \caption{Results of gain measurements performed in liquid xenon for the full set of tested PMTs. The stacked histograms correspond to all qualified PMTs (darker shaded blue) and those returned to the manufacturer for replacement (lighter shaded blue).
    Left: Distribution of amplification gains at a bias voltage of 1.5\,kV.
    The gain distribution for the XENON1T PMTs is shown in red (data from~\cite{Barrow:2016doe}).
    Right: Distribution of supply voltages  corresponding to a gain of $5\times10^6$.
    }
    \label{fig:gain1500}
\end{figure}
Figure \ref{fig:gain1500} shows the resulting distributions for the gain measurements performed in liquid xenon. The average gain at a supply voltage of 1.5\,kV for the qualified PMTs is $8.4\times10^6$ with a standard deviation of $2.3\times10^6$. Compared to the distribution of gains from the XENON1T PMT testing campaign, with an average of $5.3\times10^6$ and standard deviation of $2.1\times10^6$~\cite{Barrow:2016doe}, this constitutes a substantial improvement.
The mean and standard deviation of supply voltages corresponding to a gain of $5\times10^6$ are 1.41\,kV and 0.04\,kV for the set of qualified PMTs.

\begin{figure}[h!]
    \includegraphics{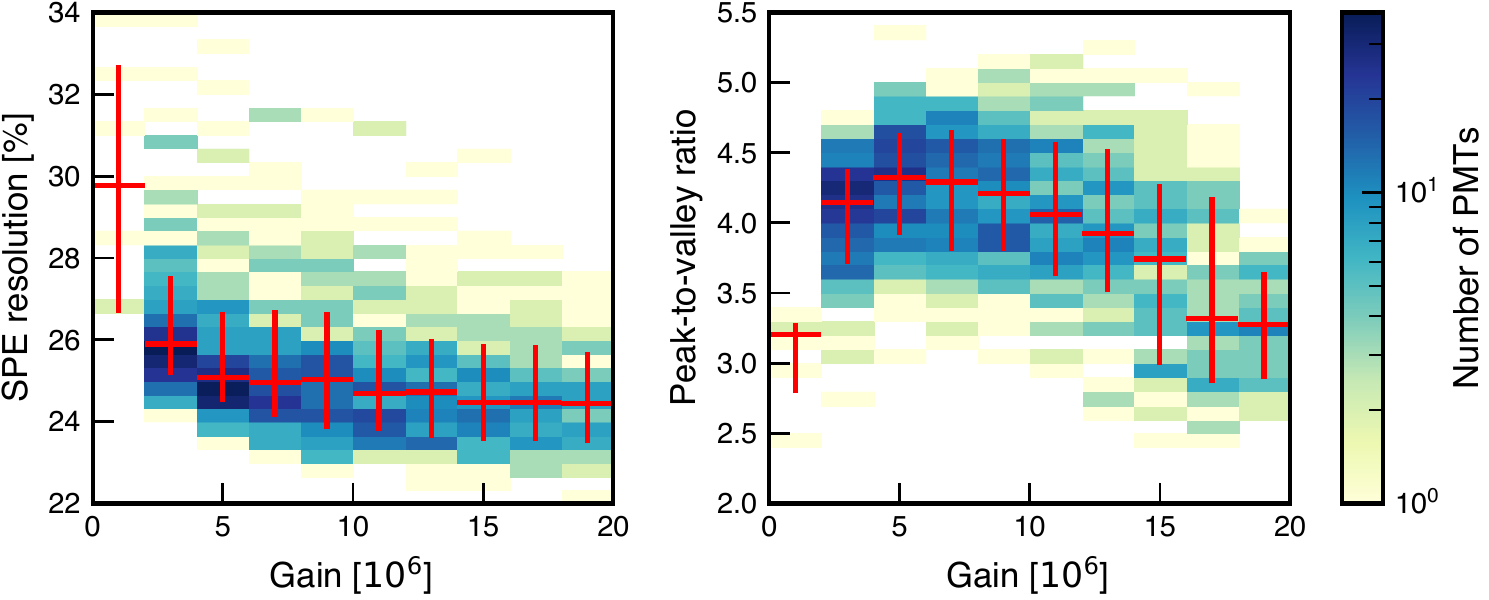}
    \caption{Dependence of the SPE resolution (left) and the peak-to-valley ratio (right) on the gain, measured for the subset of 161 PMTs tested at MPIK. Each PMT's charge spectrum is measured at seven different supply voltages. Red markers indicate the median and the 0.15 to 0.85 quantile range of the distributions in each gain bin. Both the SPE resolution and the peak-to-valley ratio plateau at a gain value of around $5\times10^6$. }
    \label{fig:speandpv}
\end{figure}

We define the SPE resolution as the ratio of the standard deviation and the mean of the fitted Gaussian SPE peak. The peak-to-valley ratio is defined as the ratio between the height of the SPE peak and the height of the minimum, or `valley', between the noise component and the SPE peak.
The distribution of the SPE resolution as a function of gain is shown in the left panel of figure~\ref{fig:speandpv}. The resolution improves with increasing gain and plateaus at a value of about 25\% for gains larger than about $5\times10^6$. Similarly, the peak-to-valley ratio distribution shown in the right panel increases for gains up to $5\times10^6$, with a maximum value of about 4.3.
We conclude that operating the PMTs at gains higher than this does not further improve their response to single photons. Given the distribution in figure~\ref{fig:gain1500}~(right), it follows that most PMTs can be operated at supply voltages considerably smaller than 1.5\,kV while maintaining an excellent single-photon response. This is crucial for long-term PMT operation in the XENONnT detector as well as the suppression of voltage-dependent micro light emission (see section~\ref{sec:le}).

%\subsubsection{Timing performance}

The timing performance of the PMTs is quantified via the electron transit time and its spread. It is defined as the time between a photon's impact on the cathode and the arrival of the electron avalanche at the anode. We determined the transit time distribution of a subset of 161 PMTs by illuminating the entire PMT window at single-photon intensity and measuring the time between the LED trigger and the arrival of the PMT pulse.
We define the transit time spread, also called jitter, as the FWHM value of a fit to the transit time distribution.
The resulting transit time spread distribution of the tested PMTs has a mean of 9.2\,ns with a standard deviation of 0.5\,ns, in good agreement with the producer’s specification of a typical transit time spread of 9\,ns.
This value for the transit time spread is rather large in comparison to other PMT types. However, it is not dominant compared to other factors that impact the timing capabilities of the XENONnT TPC, such as the decay constants (around 3\,ns and 24\,ns) of the scintillation process in liquid xenon, the sampling rate of the DAQ system of 10\,ns, and the spread of light propagation time caused by scattering in the TPC of tens of nanoseconds.

\subsection{Light emission}
\label{sec:le}
It has previously been observed that R11410 PMTs can emit light, especially when operated with a particularly high voltage~\cite{Akimov:2015cta,Baudis:2013xva,Barrow:2016doe}.
This process is not fully understood, but is generally thought to involve the emission of single photons from the internal structure of the PMT.
The wavelength of these photons is such that a significant fraction of them can be observed by the light-emitting PMT itself as well as by other PMTs.
Light emission contributes to the background in dark matter searches, and can lead to dead time if it is especially intense.

\begin{figure}[h!]
    \centering
    \includegraphics{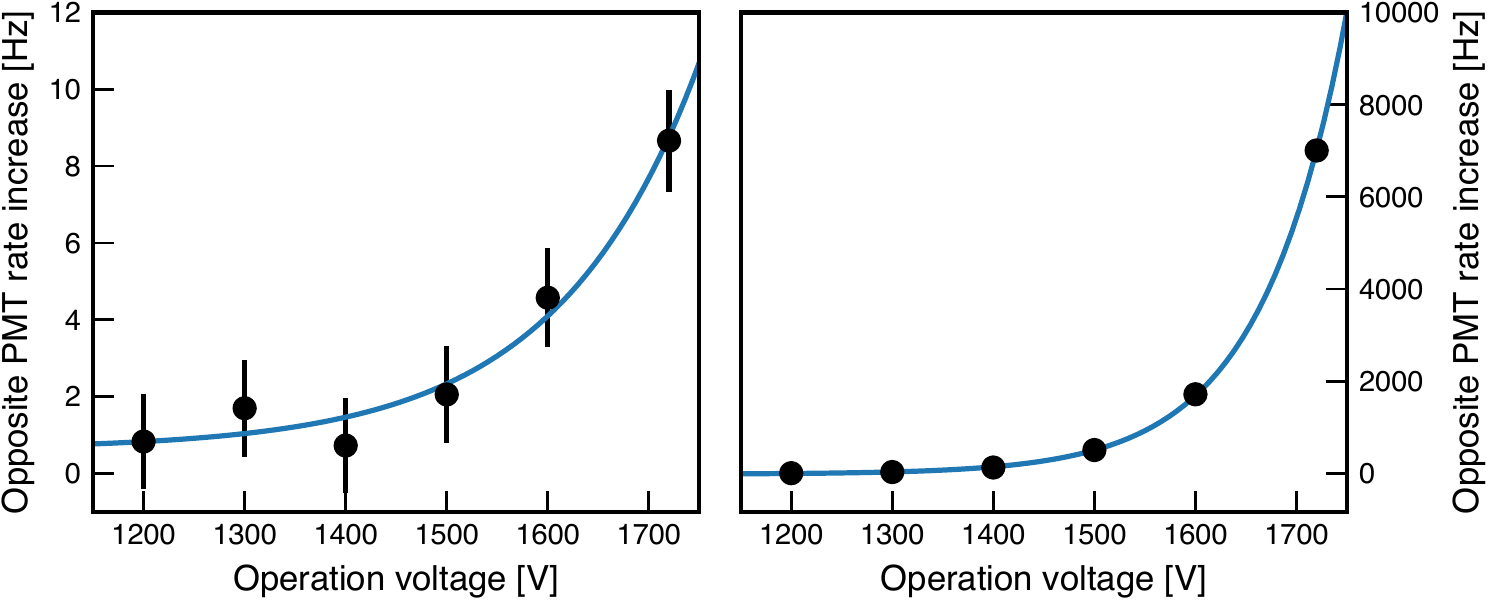}
    \caption{A measurement of the rate of micro light emission for a typical PMT (left) and a PMT with an especially high micro light emission rate (right). The increase in trigger rate of the opposite PMT is plotted against the high voltage supplied to the PMT being tested. Trigger rates are shown after subtracting the rate with the tested PMT turned off.}
    \label{fig:ule_v_dependence}
\end{figure}

The most common form, which we call micro light emission, is the apparently constant emission of low-intensity light.
This effect becomes stronger as the PMT's supply voltage is increased (see figure \ref{fig:ule_v_dependence}).
In a TPC, the random coincidence of photons originating from micro light emission, combined with single-electron signals~\cite{Edwards:2007nj}, could mimic a dark matter signal.
This combination produced a significant source of background in XENON1T, referred to as accidental coincidence~\cite{Aprile:2019dme}.

In order to identify micro light emitting PMTs, we use the same principle as in~\cite{Barrow:2016doe}.
During our tests each PMT is placed facing another window-to-window.
In the absence of LED illumination, the total rate seen by a PMT is therefore the sum of its own dark count rate and signals induced by light emitted from the opposite PMT.
If the dark count rate in the opposite PMT varies when the voltage supplied to the PMT being tested is changed, we attribute the difference in rate to micro light emission.
Figure~\ref{fig:ule_v_dependence} shows an example of such a micro light emission measurement for a typical PMT (left) and an especially bad one (right).
We characterise the amount of light emission using the difference in dark count rates of the opposite PMT, when the supply voltage is
increased from a level at which no signals are seen to an operating voltage of around 1.5\,kV--1.7\,kV.
For almost all PMTs, the rate of light emission is rather small, at a few hertz or tens of hertz when operated at 1.5\,kV.
In five of the tested PMTs we observed strong and consistent light emission, where the facing PMT saw light signals at kilohertz rates. These PMTs were immediately disqualified from being used in XENONnT.

\begin{figure}
    \centering
    \includegraphics{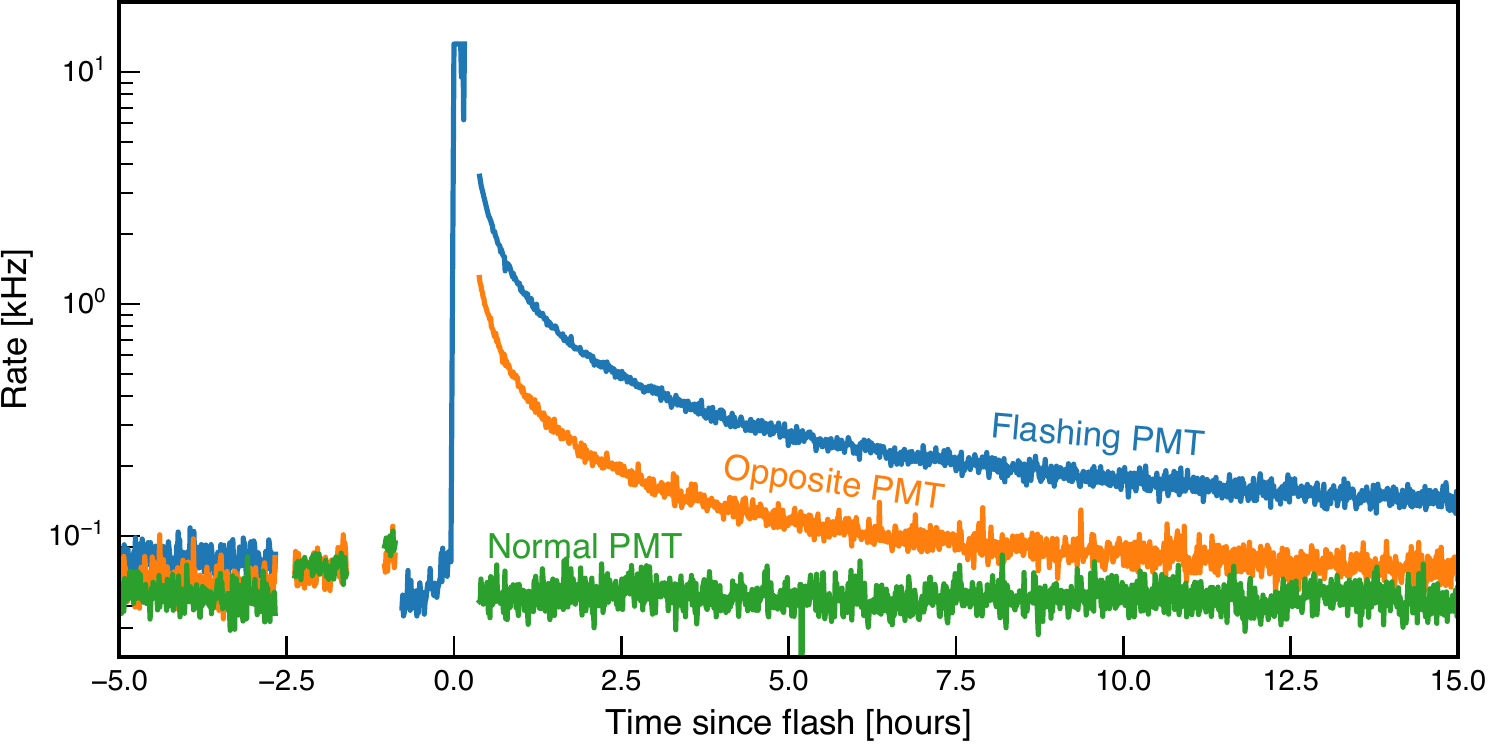}
    \caption{An example of a flashing PMT, showing the evolution of the total rate of its signals (blue line). For comparison, the PMT facing it is also shown (orange line), as is a normal PMT (green line). This flash was triggered during a micro light emission measurement.}
    \label{fig:my_label}
\end{figure}

The second category is intermittent light emission. Some PMTs generally behave normally, but occasionally show very strong light emission.
This category can be further subdivided into flashes, short bursts of extremely intense light; and much more extended periods of light emission (several hours or even days).
Flashing was first identified during the testing campaign of XENON1T PMTs~\cite{Barrow:2016doe}.
It is characterised by a suddenly appearing high rate of light emission, lasting a few seconds, followed by a gradual decrease of the count rate over several minutes or hours (figure~\ref{fig:my_label}).
This gradual decrease may be due to the usual behaviour of PMTs after exposure to bright light, when increased dark count rates are seen for some time~\cite{Akimov:2015aoa}.
During the operation of XENON1T itself, PMT flashes were identified and the surrounding time period was removed from the science data, resulting in a small (0.1\%) dead time~\cite{Aprile:2019bbb}.
Flashing was observed to be most common in PMTs with a large afterpulse rate, often induced by a high-energy deposition which creates a large light signal in the detector~\cite{Aprile:2017aty}.

\begin{figure}
    \centering
    \includegraphics{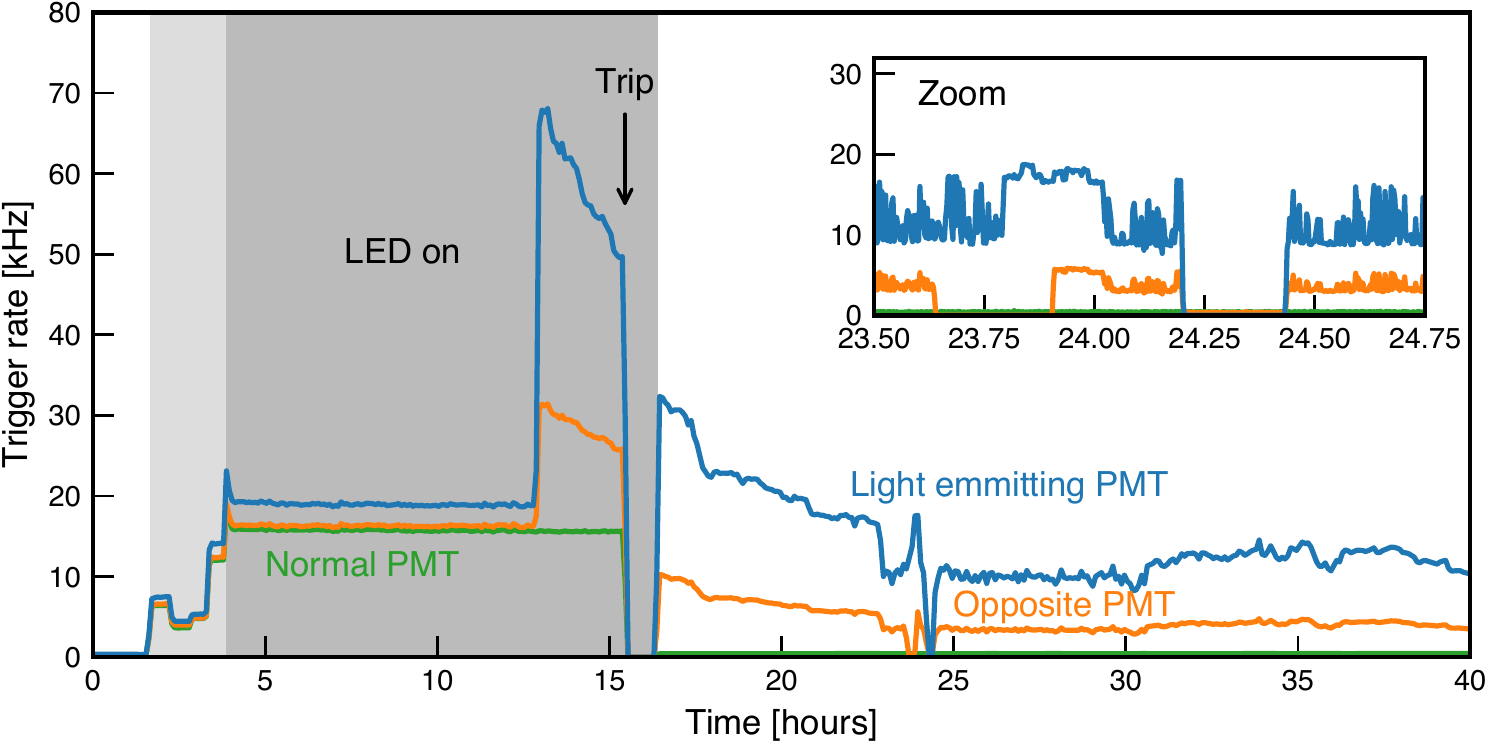}
    \caption{An example of intermittent light emission, which in this case begins during a stress test. The highest rate of triggers is recorded by the light-emitting PMT (blue line). The PMT which is facing it (orange line) also has a high trigger rate, which is clearly correlated. Another normal PMT is shown for comparison (green line) and is not affected by the light emission. The time during which the LED was turned on is shaded; including initial lower illumination (light grey) for afterpulse characterisation followed by a long period of higher intensity (darker grey). The inset shows a time when the light-emitting PMT and the one opposite were turned off in turn. When the light-emitting PMT was turned off, the trigger rate of both decreased.}
    \label{fig:stress_test_le}
\end{figure}

We introduced a so-called high-illumination stress test to the campaign. This was prompted by the experience of XENON1T, where it was observed that light emission was often triggered by a high rate of signals. This could occur while calibrating the detector or during a high-energy interaction, for example from a muon crossing the detector.
Over two periods of several hours, the LEDs were pulsed at an intensity of around $5 \times 10^5$\,PE/s, comparable to the most active calibration periods of XENON1T.
In five cases, the high illumination induced light emission, as can be seen in Figure~\ref{fig:stress_test_le}.
This generally went away after some hours and in some cases returned spontaneously, later during the test.
In total we observed intermittent light emission (lasting several hours) in six of the \ntestedtotal{} tested PMTs, which were excluded from use in XENONnT.

\subsection{Afterpulses}
\label{sec:ap}

Although photomultiplier tubes are under high vacuum, it is possible for photoelectrons to collide with and ionise residual gas particles. A resulting positively charged ion is attracted towards the photocathode, where it can cause the emission of one or more additional electrons. These will be multiplied in the same way as the photoelectron and produce a delayed pulse. This is called an afterpulse, and we refer to the electrons released from the photocathode as afterpulse electrons.
A high or increasing afterpulse rate in the PMTs can degrade the energy resolution and the accuracy of the position reconstruction of the TPC. In particularly problematic cases, it can cause other problems such as flashes (see section~\ref{sec:le}) or gradually decrease the maximum voltage at which the PMT can be operated.

For the R11410 PMT model, the time delay of an afterpulse after the main signal is proportional to the square root of the mass-to-charge ratio of the ion~\cite{Barrow:2016doe}. Therefore, the afterpulse time-delay spectrum indicates which residual gases are present in a PMT. While certain gases are expected due to the production process, others, such as xenon, should not be present. Afterpulses consistent with xenon in PMTs operated in liquid xenon clearly indicate that the tube is not leak-tight. A clear increase of the argon afterpulse rate, after operation in argon, points to the same conclusion.

In our tests, we search for afterpulses within about 5\,$\upmu$s after an LED-induced light signal and record their area in number of photoelectrons (PE) and time delay.
The afterpulses of all relevant gases lie within this time window.
We quantify the rate of afterpulses as the ratio between the number of afterpulses with an area above 2.5 PE and the number of initial photoelectrons.
This measurement is repeated multiple times, allowing us to study the time evolution of any potential leaks. 

\subsubsection{Afterpulse categorisation}
\label{sec:apcategories}

\begin{figure}[b]
\centering
\includegraphics[width=0.6\textwidth] {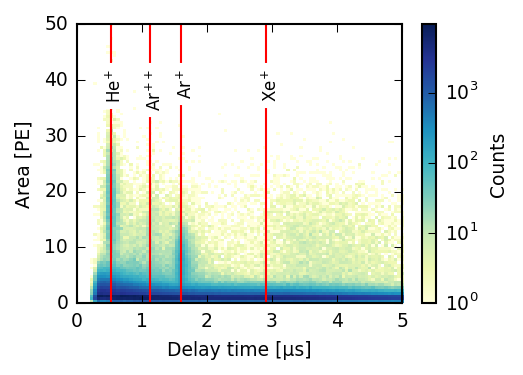}
\caption{An afterpulse spectrum of a good PMT. The expected afterpulse positions of a few common ions are indicated by the red lines.}
\label{fig:2DspectrumKB1566}
\end{figure}

Residual gases often found in the PMT vacuum include H$_2$, He, CH$_4$, N$_2$, Ar and Xe, ordered by increasing molecular mass. To remove reactive gases from the vacuum after production, the PMTs contain a strip of “cold-activated” getter material of undisclosed composition.
The noble gases He, Ar and Xe are unaffected by the getter. Of these, helium is the most commonly identified from the afterpulses it induces. Helium atoms are small enough to diffuse into the PMT. The helium contribution therefore grows slowly over time unless the PMT is stored in a helium-free atmosphere. In addition, a significant fraction of PMTs contain argon, which enters during the production process. A typical afterpulse spectrum of a normal PMT can be seen in figure~\ref{fig:2DspectrumKB1566}, containing helium and some argon.

We note that due to the similar masses of N$_2^+$ and CO$^+$, and of Ar$^+$ and CO$_2^+$, their afterpulses have similar time delays.
Some previous publications have attributed afterpulses with these delays to CO$^+$ and CO$_2^+$~\cite{Lung:2012pi}.
Our afterpulse measurements cannot distinguish between these two possibilities.
We therefore always refer to N$_2^+$ and Ar$^+$ in this manuscript.

In addition to ion afterpulses, every PMT has a contribution at low areas, whose rate drops exponentially with the time delay. This contribution comes from dark pulses and single electrons of unclear origin~\cite{Hamamatsu:handbook}.
Finally, all PMTs have a broad cluster of afterpulses of about 10\,PE area around 4\,$\upmu$s. The origin of these is not yet known.

PMTs with increased afterpulse rates can be divided into two categories.
The first has a high overall rate of afterpulses which we attribute to nitrogen contamination of the vacuum.
This may be due to an air leak or the release of a large amount of nitrogen trapped within the PMT.
The second category has an increasing rate of xenon or argon afterpulses, due to a leak during operation in these media.
These leaks may open while the PMT is being cooled down, possibly due to thermal stress, although we take care to keep temperature changes below 2\,K/min.

\begin{figure}[t]
\centering
\includegraphics[width=1.0\textwidth] {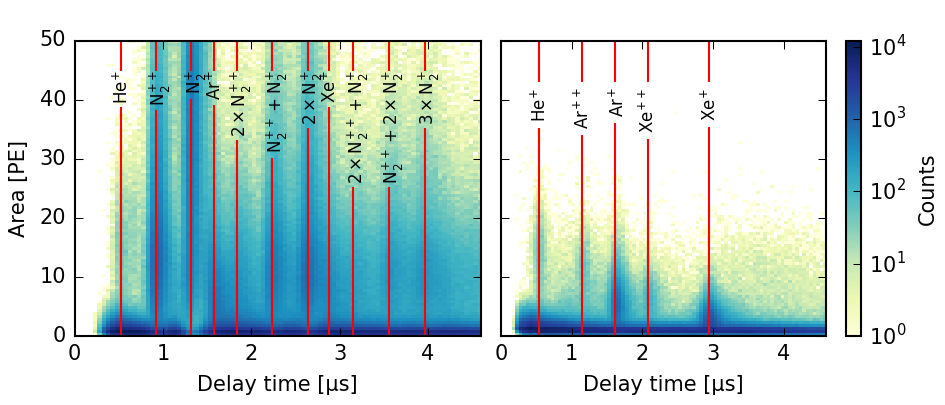}
\caption{The afterpulse spectrum in the left plot shows large afterpulses resulting from a high N$_2$ content. The 2.3\% total afterpulse rate is one of the highest we measured. The notation $2\times$N$_2^+$ refers to secondary nitrogen afterpulses after primary nitrogen afterpulses; $3\times$N$_2^+$ refers to tertiary afterpulses. The right plot shows an Xe$^+$ and an Xe$^{++}$ afterpulse peak, which is a clear indication of a leak. The Xe$^+$ afterpulse rate is 0.1\%.}
\label{fig:AP2_3}
\end{figure}

One example of the first type can be seen in figure~\ref{fig:AP2_3} (left). These PMTs have large N$_2^{+}$ and N$_2^{++}$ lines. In addition, if the gas content in the PMT is high, electrons released from the photocathode during an afterpulse have a significant probability of interacting with gas molecules before reaching the first dynode, thereby creating their own, secondary afterpulses.
Another characteristic of PMTs with high afterpulse rates is the large area of their afterpulses, often above 50\,PE. This is caused by pile-up of several simultaneous afterpulses.

Figure~\ref{fig:AP2_3} (right) shows a PMT with xenon afterpulses. Often, the Xe$^{++}$ line also becomes visible after a longer testing period. This indicates a leak, since the xenon line is not present during the first measurement in vacuum.
Some PMTs' afterpulse rates grow continuously, while the xenon or argon content in other PMTs increases in steps, for example while being cooled down.
We also observed that, in the latter case, the leak only opens during some cooling cycles, while in others the afterpulse rate stays constant. For this reason, almost all PMTs underwent at least two cooling cycles.

Of all \ntestedtotal{} measured PMTs during the XENONnT testing campaign, seven PMTs were rejected due to high overall afterpulse rates, while six were rejected due to growing rates of xenon (five PMTs) or argon (one PMT) afterpulses.
In addition, two PMTs could not be stably operated at a reasonable voltage, which we attribute to severe vacuum degradation.

\subsubsection{Afterpulse-induced light emission}
\begin{figure}[b]
\centering
\includegraphics[width=1.0\textwidth]{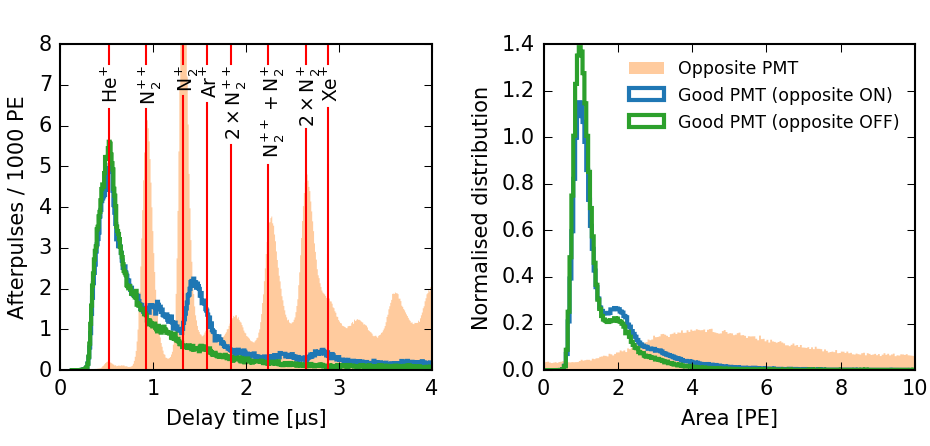}
\caption{The left plot shows the afterpulse spectra of a typical good PMT (blue) opposite a PMT which has high afterpulse rates (orange). Once the opposite PMT is switched off, the afterpulse-like lines in the good PMT disappear (green). In the right plot the normalised area distributions of the same afterpulses can be seen. When the opposite PMT is turned on, more afterpulses are visible with an area around 2\,PE, which is much lower than the area of real afterpulses.}
\label{fig:oppositePMTs_combined}
\end{figure}

In our tests, the PMTs are arranged in pairs facing each other at a distance of a few millimetres, to measure light emission (see section~\ref{sec:le}). During the afterpulse measurements, a correlation was observed between the afterpulse spectra of opposite PMTs.
If one of the two PMTs has a high rate of afterpulses, the opposite PMT shows pulses simultaneously.
However, these pulses have a much smaller area (less than 5\,PE).
After turning off the PMT with the high afterpulse rate, these unusual afterpulse-like lines in the opposite PMT disappear, as seen in figure~\ref{fig:oppositePMTs_combined} (left). 
Therefore, we infer that in PMTs with high afterpulse rates, photons are emitted in coincidence with the afterpulse. These photons are then detected by the opposite PMT. A possible explanation could be excitation processes, for example during the interaction of the ions with the photocathode material or the PMT window.
Alternatively, this light could be emitted during the collision of electrons with residual gas.
In this case, we would only see the light for secondary afterpulses, since for primary afterpulses the light would be emitted in coincidence with the LED signal.

The right panel of figure~\ref{fig:oppositePMTs_combined} shows the area spectrum of the good PMT when the opposite PMT is turned on (blue) and off (green). While the afterpulses of the leaky PMT (orange) have a typical area of 5\,PE with contributions up to very high values (pile-up effect), the distribution of afterpulse-like lines (blue) peaks at a much lower area and is incompatible with real afterpulses. It is very similar to the area distribution when the opposite PMT is turned off.

We can conclude that these afterpulse-like lines are not real afterpulses but individual photons which were emitted by the opposite PMTs.

\subsubsection{Leak detection with argon and nitrogen gas}
\label{sec:ap_ar}

\begin{figure}[h]
\centering
\includegraphics{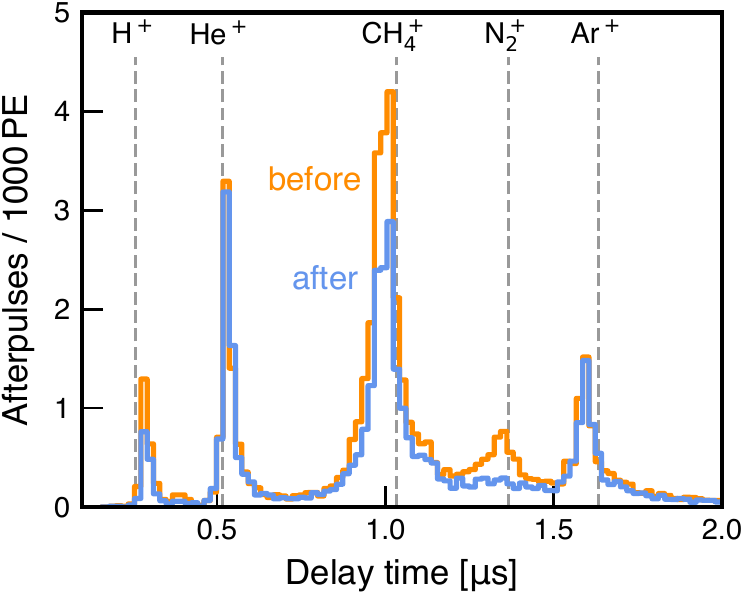}
\caption{Comparison of two afterpulse timing spectra of the same PMT, acquired before (orange) and after (blue) two cooling cycles in gaseous argon. The afterpulse lines corresponding to hydrogen H$^+$, methane CH$_4^+$ and nitrogen N$_2^+$ are reduced after the exposure to cryogenic temperatures.}
\label{fig:APions}
\end{figure}

As described in section \ref{sec:apcategories}, many non-inert trace gases are removed from the PMT vacuum by a built-in getter.
Figure~\ref{fig:APions} shows two afterpulse time spectra of the same PMT, measured at room temperature before (orange) and after (blue) the PMT underwent two cooling cycles. A decrease in the H$^+$, CH$_4^+$ and N$_2^+$ afterpulse rates is visible after the PMT was cooled down, while the contributions from the noble gases He$^+$ and Ar$^+$ remain the same.
The instability in time of the rate of nitrogen afterpulses was the reason for the low sensitivity to PMT leaks in the XENON1T testing campaign, where almost all PMTs were tested only in cold nitrogen gas.
To avoid this, we introduced the use of argon gas for the XENONnT testing campaign to monitor possible vacuum degradation in the PMTs.
One PMT showed a clear increase in the rate of argon afterpulses after two cooling cycles in argon gas.
However, the fact that most PMTs already contain a trace amount of argon (see section \ref{sec:Arleak}) makes it harder to identify an increase. We therefore also tested all PMTs in liquid and cold gaseous xenon for several weeks.

\subsubsection{Argon afterpulses as a predictor of leaks}
\label{sec:Arleak}

\begin{figure}
    \centering
    \includegraphics{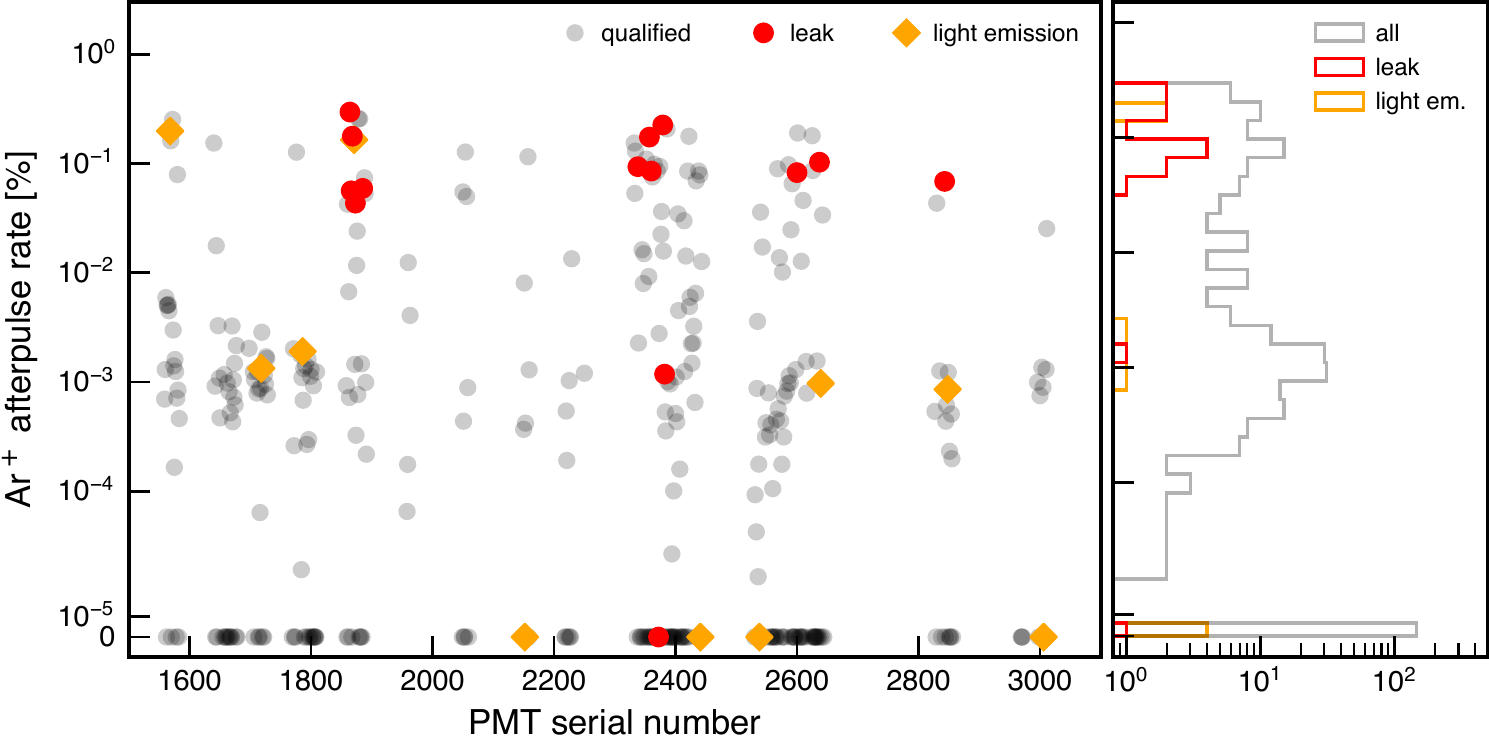}
    \caption{Argon gas amount for all tested PMTs.
    Coloured markers indicate PMTs that showed any kind of vacuum degradation (red circles) or light emission (orange diamonds). The right panel shows the histogrammed projection of the argon afterpulse rates for all PMTs as well as for PMTs that showed leaks or light emission. There is a visible correlation between the amount of argon gas in a PMT and its potential for vacuum degradation.}
    \label{fig:arvsleak}
\end{figure}

A significant fraction of PMTs contain residual argon gas that is already present on delivery, likely originating from the production process.
During the operation of XENON1T, it was observed that PMTs with a large argon afterpulse rate are likelier to develop xenon afterpulses~\cite{Rauch:2017phd}.
In figure~\ref{fig:arvsleak}, this correlation is visualised for the \ntestedtotal{} new PMTs. 
The fraction of PMTs with a detectable amount of argon is 60\%.
Of these, 20\% of PMTs have an argon afterpulse rate above 0.01\%. Twelve PMTs with an identified leak (out of a total of fifteen PMTs with leaks) belong to this subset. This corresponds to a 17\% failure rate due to vacuum degradation amongst PMTs with a high argon afterpulse rate.
In contrast, the failure rate for PMTs with an argon afterpulse rate below 0.01\% is only 1\%.
The correlation does not extend to issues related to light emission; here the failure rate is 3\% for PMTs with argon afterpulse rates both above and below 0.01\%.

For the qualified PMTs, the amount of argon gas in the tubes can indicate an increased possibility of vacuum degradation in the future. This is considered for the final selection of PMTs for the XENONnT detector as well as their arrangement.

\section{Discussion}
\label{sec:discuss}

The XENONnT PMT testing campaign was improved in several ways compared to that of XENON1T.
Most importantly, we tested all PMTs in liquid and cold gaseous xenon for several weeks in order to reliably identify vacuum degradation. 
For XENON1T, only 14\% of the PMTs were tested in this way, while all were tested in cold nitrogen gas.

The gain of the qualified new XENONnT PMTs is $(8.3\pm2.3)\times10^6$ (at an applied voltage of 1.5\,kV). This is substantially higher than the $(5.3\pm 2.1)\times 10^6$ measured during the XENON1T testing campaign, meaning we can operate the new PMTs at a lower voltage. Since problems like light emission are correlated to the supply voltage, this is expected to lower their dark count rate and improve long-term stability.

One of the main goals of our tests is the identification of various types of light emission. The vast majority of the PMTs show no micro light emission above 10\,Hz at 1.5\,kV. While some show light emission slightly above 10\,Hz, only five out of 368 tested PMTs were disqualified due to micro light emission at kilohertz rates. 
We performed a specific test to search for a new form of light emission, occurring intermittently. 
We identified six PMTs with this problem. However, due to the random nature of intermittent light emission, it is possible that some affected PMTs were not symptomatic during our tests and therefore not disqualified.

Afterpulses in PMTs can originate from residual atoms and molecules in the vacuum. In PMTs with leaks, the afterpulse rate increases over time, which can lead to problems including flashes. Two different afterpulse-related problems were detected during the testing campaign:
very large afterpulse rates which we attribute to nitrogen, seen in seven PMTs, and an increasing afterpulse contribution from xenon or argon atoms due to a small leak. Six PMTs were disqualified due to this latter problem. Moreover, two PMTs could not be operated at reasonable supply voltages, most likely due to severe vacuum degradation.
Furthermore, we observed afterpulse-like signals in good PMTs which are facing PMTs with high afterpulse rates, indicating that light can be emitted during the afterpulse process.

PMTs with vacuum degradation or light emission were disqualified from use in XENONnT. The number of disqualified PMTs, broken down by category, is summarised in table~\ref{tab:summary}. In total, 26 of 368 PMTs (7\%) were disqualified. This is much lower than the 22\% disqualified for XENON1T, despite more rigorous testing. This shows that the overall quality and leak-tightness of the PMTs is significantly better and the measures Hamamatsu took to reduce light emission were effective.

\begin{table}[h!]
  \begin{center}
    \begin{tabular}{l c r} \toprule
      \textbf{Type of problem} & \multicolumn{2}{c}{\textbf{Disqualified PMTs}} \\
      \midrule
      Air leak & 7 & 1.9\% \\
      Xe/Ar leak & 6 & 1.6\% \\
      Non-operable & 2 & 0.5\%\\
      Total vacuum degradation & 15 & 4.1\%\\ \midrule
      Micro light emission & 5 & 1.4\% \\
      Intermittent light emission & 6 & 1.6\%\\
      Total light emission & 11 & 3.0\%\\ \midrule
      Total disqualified & 26 & 7.0\%\\
      Total tested & \ntestedtotal{} & \\ \bottomrule
    \end{tabular}
    \caption{Summary of the issues which disqualified PMTs from use in XENONnT.}
    \label{tab:summary}
  \end{center}
\end{table}

\begin{figure}[ht]
\centering
\includegraphics{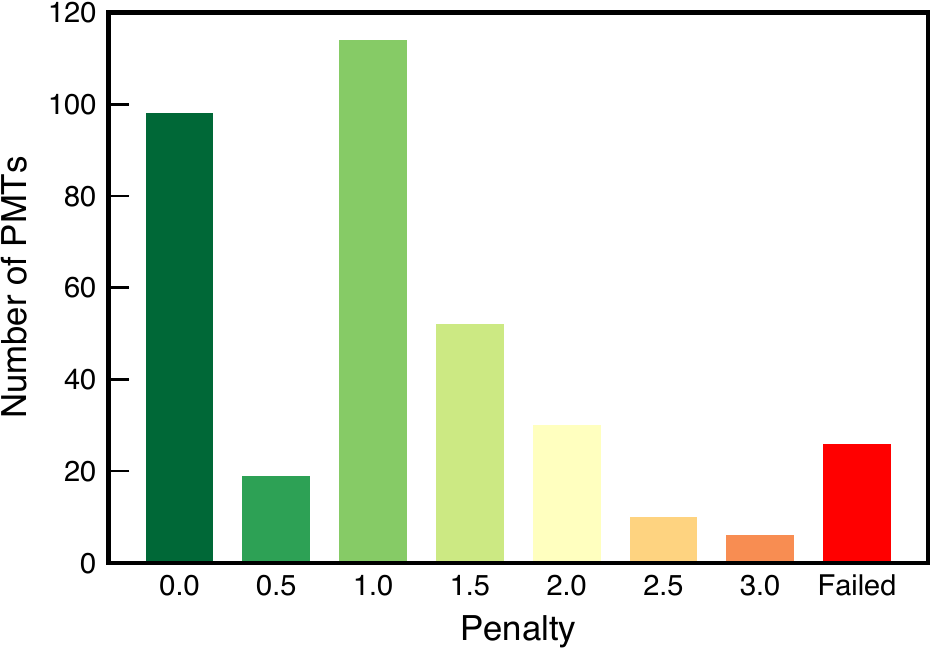}
\caption{Distribution of penalty points for all qualified PMTs.}
\label{fig:penalty_points}
\end{figure}

We classified the remaining PMTs on the basis of their test results. For this, a quantitative penalty point system was introduced.
As shown in section~\ref{sec:Arleak}, an unusually high rate of argon afterpulses can indicate an increased probability of developing a leak. Therefore, such PMTs were assigned 1.0 to 2.5 penalty points, depending on their argon afterpulse rate. Micro light emitting PMTs which were not bad enough to be disqualified were also assigned penalty points (0.5 or 1.0 depending on the severity).
Further points were awarded for low gain and extended storage time in air. Finally, since several PMTs with issues were often manufactured at a similar time, we penalised others produced in those periods.
The resulting penalty point distribution, shown in figure~\ref{fig:penalty_points}, was used to select the best PMTs for the XENONnT detector.

All the new measures -- an improved PMT in terms of micro light emission, improved and new testing facilities, tests of all PMTs in liquid and gas xenon, and new testing methods -- result in significantly more effective quality assurance for XENONnT PMTs.

\section*{Acknowledgements}

We thank Hamamatsu for the fruitful collaboration and production of the PMTs used in these studies.
We gratefully acknowledge the support of the Knut and Alice Wallenberg Foundation, the Swedish Research Council, the Max Planck Society, the ISF (1295/18) and the Weizmann Institute of Science, the Swiss National Science Foundation under Grant No. 200020-188716, the European Unions Horizon 2020 research and innovation programme under the Marie Sklodowska-Curie grant agreements No. 690575 and No. 674896, and the European Research Council (ERC) under the European Union's Horizon 2020 research and innovation programme, grant agreement No. 742789 ({\sl Xenoscope}).

\bibliography{main}

\end{document}